\documentclass[aps,pra,twocolumn]{revtex4}
\usepackage{amsfonts}
\usepackage{amssymb}
\usepackage{color}
\usepackage{amsmath}
\usepackage{graphicx}
\usepackage{subfigure}
\usepackage{txfonts}
\usepackage{bm}
\usepackage{ulem}
\usepackage[colorlinks,linkcolor=blue,citecolor=blue]{hyperref}

\begin{document}

\title{Dissipation induced information scrambling in a collision model}

\author{Yan Li, Xingli Li, Jiasen Jin}
\email{jsjin@dlut.edu.cn}
\affiliation{School of Physics, Dalian University of Technology, Dalian 116024, China}

\begin{abstract}
In this paper, we present a collision model to stroboscopically simulate the dynamics of information in dissipative systems. In particular, an all-optical scheme is proposed to investigate the information scrambling of bosonic systems with Gaussian environmental states. By varying the states of environments, we find that in the presence of dissipation the transient tripartite mutual information of system modes may show negative value signaling the appearance of information scrambling. We also find that dynamical indivisibility based non-Markovianity play dual roles in affecting the dynamics of information.
\end{abstract}
\date{\today}
\maketitle

\section{Introduction}

In closed quantum systems, the locally encoded information can spread into the nonlocal degree of freedom under the unitary transformation, which is referred to as {\it information scrambling}. The occurrence of information scrambling in the system means that the information of the initial state can not be completely accessed by any local operator after time evolution, which also hints the information delocalization in quantum many-body systems. A quantum system in which the information becomes scrambled can be viewed as a quantum chaotic system: a local operator grows under the time evolution to have large commutators with almost all other operators in the system \cite{hosur2016}. The most efficient scrambled system in nature is the block hole. Meanwhile the Sachdev-Ye-Kitaev (SYK) model\cite{SYmodel}, a toy model of low-dimensional quantum black holes, is also well-known in the field of condensed matter.

A well-known indicator for information scrambling is the out-of-time-order correlator (OTOC) \cite{BraianSwingle2016,Duanluming2019,Joserau2019,KALandsman}, which measures the overlap of operators in the dynamics and is intimately  related to the Lyapunov exponent. The OTOC is not only widely used to investigate the quantum chaos \cite{JHEP082016,PRL1152015,JHEP042016,JHEP052017,JHEP042017}, but also plays a dramatic role in characterizing phase transition \cite{HuitaoShen2017,MarkusHeyl2018,QianWang2019,PRL1232019CB,PRL1232019Subhayan2019,SoonwonChoi2020,RohitKumarShukla2021} and many-body localization \cite{YHuang2016,RFan2017,XChen2016,RQHe2017,BSwingle2017}. Although the experimental realization of the inverse-time evolution is challenging, the OTOC is still experimentally investigated in trapped ions \cite{MGarttner2017} and nuclear magnetic resonance quantum simulators \cite{JLiRFan2017}.

The tripartite mutual information (TMI) provides an alternative path to study the information scrambling without the inverse-time evolution \cite{Eiki2018,Oskar2019,HuitaoShen2020,DarvinWanishch2021}. In Ref. \cite{hosur2016}, Hosur {\it et al.} introduced a map from a unitary quantum channel to a state in a doubled Hilbert space, through which the OTOC and TMI are connected. The information scrambling can be characterized by the negative value of TMI. It has already been used to study the weak and strong thermalization \cite{zhengHangSun2021}, information delocalization \cite{WanischPRA2021} and quantum frustration \cite{MatsudaPRE2000,MatsudaPhysica2001} in past few decades. Although TMI has the advantage of being operator-independent in witnessing the information scrambling, the exponentially increasing dimension of Hilbert space still limits the further investigations. Meanwhile, an absolutely closed or isolated system does not exist in reality, imperfect experimental conditions and measurements always induce the noisy environments. As a consequence, the studies on the information scrambling in noisy systems have received much attention in recent years \cite{Michael2018PRB,Brian2018PRBA,Syzranov2018PRB,Zhang2019PRB,yl2020PRA,Bibek2021PRB,Dominguez2021PRA}, and the investigations on the dynamics of information in dissipative systems are still desired.

In this paper, we utilize the {\it collision model} (CM) to tackle those puzzles in open systems. By means of the CM, we can simulate considerable many-body interactions and can also structure different noisy environments: Markovian and non-Markovian cases to study the non-Markovianity effect on the information scrambling. The idea of CM is to represent the system with a particle and the environment with an ensemble of identical particles. The continuous interactions between the system and its environment are thus simulated by a sequence of collision processes. If the system always collides with a fresh environmental particle at each step, the information of the system irreversibly flows to the environment which means that the dynamics of the system is Markovian. On the other hand, if the system collides with the environmental particles who contain the history of the information, the dynamics are considered to be non-Markovian. Apart from the works for investigating the non-Markovian dynamics \cite{VScarani2002,FCiccarelloPRA2013,FCiccarelloPS2013,mccloskey2014,bernardes2014,JiasenJin2015,Bcakmak2017,bernardes2017,JiasenJin2018,ZhongXiaoMan2018,SJWhalen2019,Rolando2021}, CM has also been used to study the quantum synchronization \cite{GKarpat2019}, quantum steering \cite{KBeyer2018}, multipartite dynamics \cite{SLorenzoPRA2017}, multipartite entanglement generation \cite{Baris2019},  quantum friction \cite{grimmer2019} and thermodynamics \cite{strasberg2017,shao2018,cusumano2018,man2019,Rodrigues2019}, in particular for the quantum thermometry \cite{alves2022,landi2021}. A comprehensive panorama of the studies on CM can be found in Ref. \cite{ciccarello2021}. Recently, the experimental realization of CM for non-Markovian dynamics in all-optical system has been reported \cite{Cuevas2019}.

\begin{figure*}[!htp]
  \includegraphics[width=1.0\linewidth]{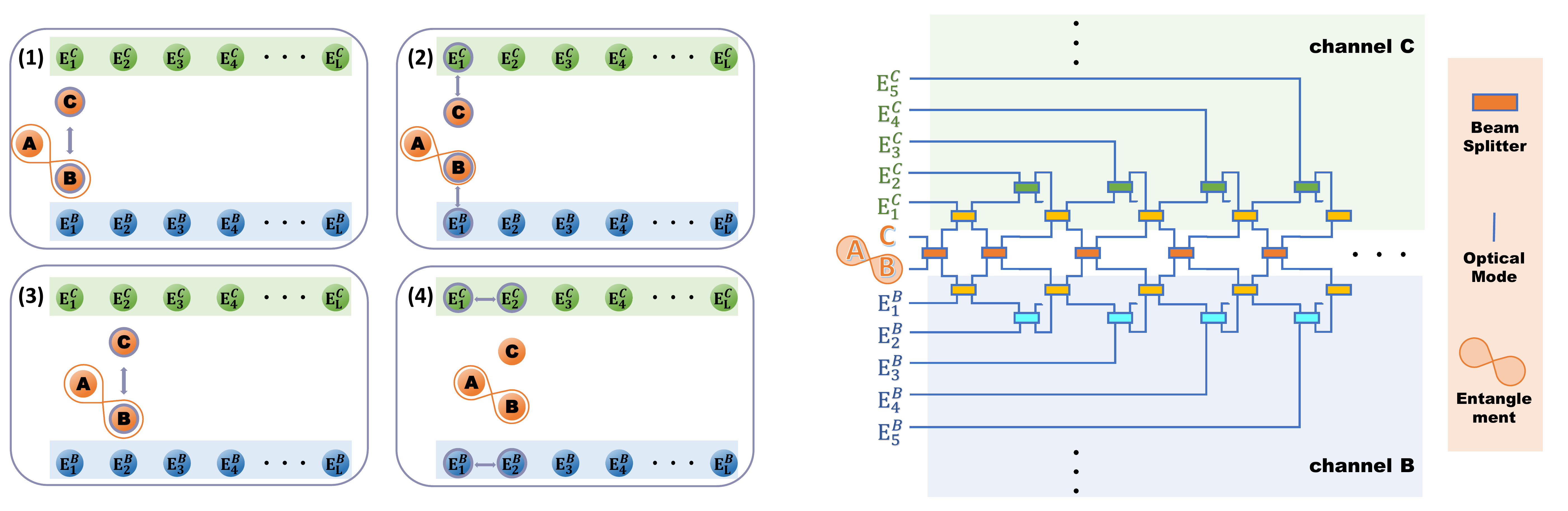}
  \caption{ The left panel is the collision route and the right one is the pictorial illustration of our model. There are two kinds of optical modes, one is system part consisting of $A$, $B$ and $C$ of which $A$ and $B$ are entangled and $A$ will not involve in the evolution. And the other is environment part: $E_{j}^{B}$ and $E_{j}^{C}$ which are in the dissipative channel $B$ and channel $C$, respectively. As the collision route shows, the whole process is divided into four parts, and the dynamic process repeats collisions (4),(2),(3) after the first three collisions (1)-(3) occurs. (1) The collision between $B$ and $C$. (2) The collision between system part and their own environment part. (3) The collision between $B$ and $C$. (4) The collision between environment part. The collision can be realized by BS, as the pictorial illustration shows the orange BS is utilized to mix the two system modes $B$ and $C$. The yellow BS is to mix the system mode and environmental mode. As for the blue and green BSs are represented the mixture of two environmental modes in their each dissipative channel $B$ and $C$.}
  \label{Fig_Model}
\end{figure*}

Here we propose an experimental feasible scheme to implement the CM in an all-optical network. This allows us to simulate the information dynamics of the continuous-variable system. The all-optical network is consisted of linear optical elements such as the beam splitters (BSs). In such scheme, the system and environmental particles are represented by the optical modes in different optical paths. We consider a joint tripartite system composed of an auxiliary mode $A$ and two system modes $B$ and $C$. Moreover, $B$ and $C$ are subjected to individual dissipative channel while mode $A$ is isolated. The initial information is encoded locally in mode $B$. The interactions between different modes are realized through the BSs. By modulating the transmissivities (or reflectivities) of the BSs, the dissipative channels of the system modes can be tuned from the Markovian to the non-Markovian cases. All the input modes are restricted to be the Gaussian states. The non-Markovianity can be quantified by the degree of the violation of dynamical divisibility \cite{plenio2010,GTorre2015}. We adopt the TMI as the measure of the information scrambling which can be calculated through the symplectic eigenvalues of the so-called covariance matrix \cite{ASerafini2004,ASerafini2006}.

In the presence of dissipation, we find that before the information completely lost to the environment, the local information at the initial time will spread into the non-local Hilbert space during the evolution and can not be collected by local operators. This phenomenon is referred as information scrambling which can be indicated by the negative value of TMI, and the physical significance of the negative value of TMI is that the information can be non-locally stored in different subsystems simultaneously. Then we focus on Markovian (non-Markovian) dynamics, we find that the non-Markovianity can indeed affect the time duration of information scrambling, but it is not the key factor for the emergence of information scrambling. We expect that our work may provide a sufficient theoretical support for the experimental studies on the information delocalization in dissipative system. It should be noted that a CM study on the information scrambling in a many-body system with local dissipation has been proposed in Ref. \cite{yl2020PRA}. It is shown that the disorders, nonuniform interactions and interplay between the nearest-neighboring interactions and local dissipation contribute to the information scrambling in a many-body system. In contrast, in this work we concentrate on the roles of dissipations for the information scrambling in the absence of many-body effects.

The paper is organized as follows. In Sec. \ref{model}, we explain the idea of our CM and the linear optical setup. The mathematical descriptions of the our model are also introduced. In Sec. \ref{Scheme}, we give the derivation for the degree of the non-Markovianity and TMI based on our CM. In particular, we also present the regime of Markovian and non-Markovian channels in the parameter space. In Sec. \ref{resultsanddiscussion} we show the dynamics of the TMI with different initial states as well as the temperature of environments for both Markovian and non-Markovian cases. We summarize in Sec. \ref{summary}.

\section{Collision model}
\label{model}

The considered system is comprised of three parts, labeled by $A$, $B$ and $C$ which are considered as the auxiliary mode and system modes, respectively. The auxiliary mode $A$ is isolated while the system modes $B$ and $C$ interact with each other and with their individual environments $E^B$ and $E^C$. The dynamics of modes $B$ and $C$ can be investigated through the CM in which each system modes is represented by a particle and the environments are represented by ensembles of identical particles. We label the $j$-th environmental particle of the dissipative channel as $E_j^{B(C)}$ with $j$ = $1,2,...,L-1$. The intra-system interaction is simulated by the collisions between particles $B$ and $C$ while the system-environment interactions are represented by the collisions between the corresponding system and environmental particles. As shown in Fig. \ref{Fig_Model}, our CM works through the following collisions,
\begin{description}
  \item[(1)] The collision between $B$ and $C$ takes place.
  \item[(2)] $B$ and $C$ collide with $j$-th environmental particles $E_j^{B(C)}$ individually.
  \item[(3)] $B$ and $C$ collide with each other again.
  \item[(4)] The environmental particles $E_{j}^{B(C)}$ interact with $(j+1)$-th mode $E_{j+1}^{B(C)}$.
\end{description}
On the basis of the first three collisions (1)-(3), and then repeating collisions (4),(2),(3), the continuous dissipative dynamics of the system can be stroboscopically simulated by this sequence of collisions.

The considered CM can be realized in an all-optical network as shown in Fig. \ref{Fig_Model}. In the all-optical scheme, the system and environmental particles can be described by the different optical modes. The interaction between arbitrary two particles can be realized by mixing the corresponding input modes through BS. The input and output modes are linked by the so-called {\it scattering matrix} in the following way,
\begin{eqnarray}
\begin{pmatrix}
 \hat{a}_1^{\text{out}} \\[0.3em]
 \hat{a}_2^{\text{out}} \\
\end{pmatrix} = \mathcal{S} \begin{pmatrix}
\hat{a}_1^{\text{in}}\\[0.3em]
\hat{a}_2^{\text{in}}\\
\end{pmatrix} ,
\label{Scattering matrix1}
\end{eqnarray}
where $\hat{a}$ and $\hat{a}^\dagger$ are the annihilation and creation operators of bosonic mode and $\mathcal{S}$ is the scattering matrix, which is given by the follows,
\begin{eqnarray}
\mathcal{S} =
\begin{pmatrix}
r & t \\
-t & r \\
\end{pmatrix}.
\label{Scattering matrix2}
\end{eqnarray}
where $r=\sin \theta$ and $t = \cos \theta$ are the reflectivity and transmissivity of the BS, and $\theta \in [0,\pi/2]$ is the tunable parameter. Notice that the reflectivity and transmissivity satisfy $r^2+t^2=1$. Basically, there are three types of collision in our CM: (i) The collision between $B$ and $C$, the scattering matrix is labeled as $\mathcal{S}_{SS}$ and the parameter is $\theta_{ss}$. (ii) The collision between $B$ ($C$) and its environmental modes. We label the scattering matrix as $\mathcal{S}_{SE_j}$ and the parameter as $\theta_{se_j}$. We set the strengths of all the collisions associated to $\mathcal{S}_{SE_j}$ to be identical, that is $\theta_{se_j} = \theta_{se}$, $\forall j$ and the corresponding reflectivity $r_{se_j} = r_{se}$, $\forall j$ and transmissivity $t_{se_j} = t_{se}$ $\forall j$. (iii) The collision between environmental modes. The scattering matrix and parameter are $\mathcal{S}_{E_jE_{j+1}}$ and $\theta_{e_je_{j+1}}$. Here, again we set the strengths of all the collisions associated to $\mathcal{S}_{E_jE_{j+1}}$ to be identical, that is $\theta_{e_je_{j+1}} = \theta_{ee}$, $\forall j$ and the corresponding reflectivity $r_{e_je_{j+1}} = r_{ee}$, $\forall j$ and the transmissivity $t_{e_je_{j+1}} = t_{ee}$, $\forall j$.

In this paper, we restrict the input modes to be a subset of Gaussian states with zero first moments. The linear optical elements in the CM always preserve the Gaussianity of the input states and will not introduce the first moments. Moreover, the reason why we do not consider the coherent states as the initial states is that the entanglement is undoubtedly an important condition for observing information scrambling, however, in the evolution of our current model, the coherent states will not produce the entanglement. Meanwhile, Compared with the other states, the squeezed vacuum state as the initial state has the advantage of simplifying calculations. At this sense, the network used here to realize the CM is recognized as Gaussian channels. In order to simulate the time-evolution of the system state, we use the ``temporal" index $L$ to denote the times of collisions of modes B and C in our CM. With the number of the collisions increasing, new optical modes are involved into the scattering matrix at each step. After $L$-times collision, the input-output relation of the optical modes is $[\hat{a}^{\text{out}}_{E^B_{L-1}},\hat{a}^{\text{out}}_{E^B_{L-2}},...,\hat{a}^{\text{out}}_{B},\hat{a}^{\text{out}}_{A},\hat{a}^{\text{out}}_{C},...,\hat{a}^{\text{out}}_{E^C_{L-2}},\hat{a}^{\text{out}}_{E^C_{L-1}}]^{\text{T}} = \mathbb{S}(L)[\hat{a}^{\text{in}}_{E^B_{L-1}},\hat{a}^{\text{in}}_{E^B_{L-2}},...,\hat{a}^{\text{in}}_{B},\hat{a}^{\text{in}}_{A},\hat{a}^{\text{in}}_{C},...,\hat{a}^{\text{in}}_{E^C_{L-2}},\hat{a}^{\text{in}}_{E^C_{L-1}}]^{\text{T}}$, here the subscripts $X=A,B$ and $C$ represent the corresponding system modes and $E^{X}_{j}$ ($j=1,2,...,L-1$) denotes the $j$-th mode of the environment of system $X$. The superscript $\text{T}$ denotes the transpose of the vector (or matrix). As shown in Fig.\ref{Fig_Model}, at the first step ($L=1$), the collision only happens between the system modes, so the total scattering matrix can be constructed by $\mathbb{S}(1) = \mathcal{S}_{SS}$. In the next step, the environmental modes are involved to collide with the corresponding system modes. Thus the total scattering matrix for $L=2$ is $\mathbb{S}(2) = \mathcal{S}_{SS}\mathcal{S}_{SE_{1}}\mathbb{S}(1)$(here the subscript of $\mathcal{S}_{SE_{1}}$ denotes the scattering matrix linked the system and the first mode of the corresponding environmental mode $E_{1}^{B}$ and $E_{1}^{C}$.)  For $L > 2$, the environment-environment collision involved. As we mentioned before, one the basis of the total scattering matrix for $L=2$, be repeating the collision (4),(2),(3) in order, the total scattering matrix is given as follows
\begin{eqnarray}
\mathbb{S}(L) = \displaystyle\prod_{j=1}^{L-2}\left(\mathcal{S}_{SS}\mathcal{S}_{SE_{j+1}}\mathcal{S}_{E_{j}E_{j+1}}\right)\mathbb{S}(2).
\label{total_scattering_matrix}
\end{eqnarray}
The scattering matrices presented in Eq. (\ref{total_scattering_matrix}) are given by
\begin{eqnarray}
\mathcal{S}_{SS} =
\begin{pmatrix}
I_{L-1}&0&0&0&0\\
0&r_{ss}&0&t_{ss}&0\\
0&0&1&0&0\\
0&-t_{ss}&0&r_{ss}&0\\
0&0&0&0&I_{L-1}\\
\end{pmatrix},
\end{eqnarray}
where $I_{N}$ is the $N\times N$ identity matrix,
\begin{eqnarray}
\mathcal{S}_{SE_{j+1}}=
\begin{pmatrix}
I_{L-j-2}&0&0&0&0&0&0&0&0\\
0&r_{se}^{B}&0&-t_{se}^{B}&0&0&0&0&0\\
0&0&I_{j}&0&0&0&0&0&0\\
0&t_{se}^{B}&0&r_{se}^{B}&0&0&0&0&0\\
0&0&0&0&1&0&0&0&0\\
0&0&0&0&0&r_{se}^{C}&0&t_{se}^{C}&0\\
0&0&0&0&0&0&I_{j}&0&0\\
0&0&0&0&0&-t_{se}^{C}&0&r_{se}^{C}&0\\
0&0&0&0&0&0&0&0&I_{L-j-2}\\
\end{pmatrix},
\label{Eq:SSE}
\end{eqnarray}
where the superscript of the reflectivity (transmissivity) indicates the corresponding dissipative channel and
\begin{eqnarray}
\mathcal{S}_{E_{j}E_{j+1}}=
\begin{pmatrix}
I_{L-j-2}&0&0&0&0&0&0\\
0&r_{ee}^{B}&-t_{ee}^{B}&0&0&0&0\\
0&t_{ee}^{B}&r_{ee}^{B}&0&0&0&0\\
0&0&0&I_{2j+1}&0&0&0\\
0&0&0&0&r_{ee}^{C}&t_{ee}^{C}&0\\
0&0&0&0&-t_{ee}^{C}&r_{ee}^{C}&0\\
0&0&0&0&0&0&I_{L-j-2}\\
\end{pmatrix}.
\label{Eq:SEE}
\end{eqnarray}

\section{The formalism}
\label{Scheme}

\subsection{Characteristic function of Gaussian state}
Since the input modes are restricted to be Gaussian states and the channel is Gaussian in the CM, it is convenient to describe the state in characteristic function formalism \cite{DFWalls1994}. For a given quantum state with density matrix $\rho$, the corresponding characteristic function is given by $\chi[\lambda] = \text{tr}[\rho D(\lambda)]$, where $D(\lambda)$ is the Weyl displacement operator having the form $D(\lambda) = \exp(\lambda \hat{a}^{\dagger} - \lambda^{*}\hat{a})$.

In our model, the information is initially encoded in system mode $B$ through the entanglement between $B$ and the auxiliary mode $A$. Our goal is to investigate how the localized information spreads over the system $BC$. The entangled state of modes $A$ and $B$ is set to be the two modes squeezed vacuum state (TMSV) $|\text{TMSV}(\xi)\rangle=\hat{S}(\xi_{AB})|0\rangle_{A}|0\rangle_{B}$. The vector $|0\rangle$ represents the vacuum state and $\hat{S}(\xi_{AB})$ is the two-mode squeezing operator given as follows,
\begin{eqnarray}
\hat{S}(\xi_{AB})=\exp{\left(\frac{1}{2}\xi_{AB}^* \hat{a}_A\hat{a}_{B}-\frac{1}{2}\xi_{AB} \hat{a}_A^\dagger \hat{a}_{B}^\dagger\right)},
\end{eqnarray}
where $\xi_{AB} = r_{AB}e^{i\phi_{AB}}$ is the squeezing parameter, and $r_{AB}$ is the squeezing strength, $\phi_{AB}$ is the squeezing angle. Meanwhile, the system mode $C$ is initialized in a generic single mode squeezed vacuum state, which is expressed as $|\xi_{C}\rangle=\exp{(\frac{1}{2}\xi_C^* \hat{a}^{2}_C-\frac{1}{2}\xi_C \hat{a}_C^{\dagger2})}|0\rangle$. Therefore the joint characteristic function of modes $A$, $B$ and $C$ is given by
\begin{eqnarray}
\begin{aligned}
\chi^{\text{in}}_{ABC}\left(\vec{\mu}\right)=&\exp{\left( \frac{\mu_A\mu_B+\mu_A^*\mu_B^*}{2}\sinh{\xi_{AB}}\right)}\\
&\times\exp{\left( -\frac{|\mu_A|^2+|\mu_B|^2}{2}\cosh{\xi_{AB}}  \right)}\\
&\times \exp{\left( -\frac{|\mu_C|^2}{2}\cosh \xi_C + \frac{e^{-i\phi_c}\mu_C^2 + e^{i\phi_c}\left( \mu_C^*\right)^2}{2} \sinh \xi_C \right)},
\end{aligned}
\label{CF_sys}
\end{eqnarray}
where $\xi_C=r_Ce^{i\phi_C}$ is the squeezing parameter of the system and $\phi_C$ is the squeezing angle of $C$ mode. For simplicity, we set the squeezing parameter $\xi_{AB}$ of the TMSV state to be real.

On the other hand, each environmental mode is in the generic single-mode Gaussian state $\rho_{E_j}(n_{E_j},\xi_{E_j},\alpha_{E_j})$, where $n_{E_j}$ is the thermal mean photon number, $\xi_{E_j}=r_{E_j}e^{i\phi_{E_j}}$ is the squeezing parameter of environmental mode, $\phi_{E_j}$ is rotating angle and $\alpha_{E_j}$ is the complex displacement \cite{Marian2004}. The corresponding characteristic function is given as follows,
\begin{eqnarray}
\chi^{\text{in}}_{E_j} \left(\mu_{E_j} \right) = \exp{\left[ - X_{E_j}|\mu_{E_j}|^2 - \frac{1}{2} \left( Y_{E_j}^*\mu_{E_j}^2 + Y_{E_j}\mu_{E_j}^{*2} \right) + Z_{E_j}\mu_{E_j}^* - Z_{E_j}^*\mu_{E_j}\right]}.
\label{eq.OneModeCharacteristicFunction}
\end{eqnarray}
where $X$ ,$Y$ and $Z$ are related to the properties of the Gaussian state. The specific forms can be obtained as,
\begin{eqnarray}
X_{E_j} & = & \left(n_{E_j} + \frac{1}{2}\right)\cosh \left(2r_{E_j}\right),\cr\cr
Y_{E_j} & = & - \left( n_{E_j} + \frac{1}{2} \right) \sinh \left(2r_{E_j}\right)e^{i\phi_{E_j}} , \cr\cr
Z_{E_j} & = & \alpha_{E_j}.
\label{eq.GaussianState}
\end{eqnarray}
For concise, we have omit the notation of $B$ and $C$ in Eqs. (\ref{eq.OneModeCharacteristicFunction}) and (\ref{eq.GaussianState}).
Consequently, the total characteristic function of the system and environmental modes can be described by,
\begin{eqnarray}
\chi_{J}^{\text{in}}(\vec{\mu}) = \displaystyle\prod_{j=1}^{L-1}\chi_{j}^{\text{in}}(\vec{\mu}_{E_j^B}) \times \chi_{ABC}^{\text{in}}(\vec{\mu}_{ABC}) \times \displaystyle \prod_{j=1}^{L-1} \chi_{j}^{\text{in}}(\vec{\mu}_{E_j^C}),
\label{eq.JointCharacteristicFunction}
\end{eqnarray}
where $\vec{\mu}=[\vec{\mu}_{E^B},\vec{\mu}_{ABC},\vec{\mu}_{E^C}]$ is a vector of variables corresponding to the environmental modes $E^B$, joint system modes, and environmental modes $E^C$, respectively. The reduced characteristic function for the modes of interest can be obtained by simply setting the variables associated to the rest modes in $\vec{\mu}$ to be zero \cite{XBWang2007}. For instance, by setting $\vec{\mu}=\left[0,\cdots,0,\mu_A,\mu_B,\mu_C,0,\cdots,0\right]$ in Eq. (\ref{eq.JointCharacteristicFunction}) we could recover the characteristic function for the system modes, i.e. Eq. (\ref{CF_sys}). With the help of the scattering matrix in Eq. (\ref{total_scattering_matrix}), the input-output relation for the characteristic functions is given by
\begin{eqnarray}
\chi_{J}^{\text{out}}(\vec{\mu})=\chi_{J}^{\text{in}}(\mathbb{S}^{-1}\vec{\mu}).
\label{eq.RelationInAndOut}
\end{eqnarray}

The covariance matrix and the characteristic function are equivalent in describing the properties of Gaussian states with null vector of first moments. In particular, it is very convenient to quantify the measure of non-Markovianity of Gaussian channel and TMI in Gaussian states in terms of covariance matrix. For a single-mode characteristic function, the covariance matrix is the second moment of it and the elements are defined by
\begin{eqnarray}
\sigma_{ml} =
\frac{1}{2} \langle \hat{x}_{m}\hat{x}_{l} + \hat{x}_{m}\hat{x}_{l} \rangle - \langle\hat{x}_{m}\rangle \langle\hat{x}_{l}\rangle, (m,l= 1,2),
\label{eq.CovarianceMatrix}
\end{eqnarray}
where $\langle\cdot\rangle$ is the expectation value. We have defined $\hat{x}_{1} = (\hat{a}_{j} + \hat{a}_{j}^{\dagger})/\sqrt{2}$ and $\hat{x}_{2} = (\hat{a}_{j} - \hat{a}_{j}^{\dagger})/\sqrt{2}i$.
 The symmetrically ordered moments can be calculated through the single-mode characteristic function as follows,
 \begin{eqnarray}
\text{tr}\Big\{\rho\left[(\hat{a}_{m}^{\dagger})^{p}\hat{a}_{l}^{q}\right]_{\text{symm}}\Big\}= (-1)^{q}\frac{\partial^{p+q}}{\partial \mu_{m}^{p} \partial \mu_{l}^{\ast q}}\chi(\mu)\Big\vert_{\mu=0} .
\label{eq.OrderedMoment}
 \end{eqnarray}
For a $N$-mode Gaussian state, the corresponding covariance matrix is $2N$-dimensional. According to Eqs. (\ref{eq.RelationInAndOut})-(\ref{eq.OrderedMoment}), the covariance matrix for modes $A$, $B$ and $C$ when the environmental states are fixed is,
\begin{eqnarray}
 \sigma_{ABC} = \left(
 \begin{array}{ccc}
 \sigma_{A}&\sigma_{AB}&\sigma_{AC}\\[0.3em]
 \sigma_{AB}^{\text{T}}&\sigma_{B}&\sigma_{BC}\\[0.3em]
 \sigma_{AC}^{\text{T}}&\sigma_{BC}^{\text{T}}&\sigma_{C}\\
 \end{array}
 \right),
 \label{eq.ABCCovarianceMatrix}
 \end{eqnarray}
where $\sigma_{x}$ ($x = A,B,C,AB,AC,BC$) is a $2\times2$ matrix. The details of those covariance matrices are given in the Appendix.

\subsection{Non-Markovianity of the dissipative channel}
\label{NM}
The all-optical network can be considered as a quantum channel that evolve the input states into the output states.
A quantum channel is non-Markovian if the memory effect is present. Usually, the non-Markovianity is characterized in two routes: one is based on the information backflow in the time-evolution of a quantum system  \cite{breuer2009,laine2010} and the other is based on the degree of the violation of the divisibility of dynamical map \cite{plenio2010,GTorre2015}.
In particular, the non-zero backflow of information is shown to be a sufficient (not necessary) condition for the indivisibility of the dynamical map \cite{HaikkaPRA2011,TrapaniPRA2016}. In this work, we will adopt the measure of non-Markovianity for Gaussian channels proposed by Torre {\it et al.} in Ref. \cite{GTorre2015}.

In our CM, modes $B$ and $C$ (although they interact to each other) are subjected to individual and identical environments $E^{B}_{j}$ and $E^{C}_{j}$, respectively. Our goal is to investigate the effects of $E^{B}_{j}$ and $E^{C}_{j}$ on the dynamics of the information initially encoded in $B$. With this goal in mind, we first characterize the non-Markovianity for the dissipative channels of $C$ (the result is shown in Eq.(\ref{eq.NMequation})). Then we attached such (identical) channels to $B$ and $C$, respectively.

In order to characterize the non-Markovianity of the dissipative channel for mode $C$, we construct a scattering matrix, which is denoted by $\tilde{\mathbb{S}}_{E}(L)$, for a simplified model consisted of only mode $C$ and its environmental modes $E_j^C$. The specific form of $\tilde{\mathbb{S}}_{E}(L)$ is given as follows,
\begin{eqnarray}
\tilde{\mathbb{S}}_{E}(L) = \displaystyle\prod_{j=1}^{L-2}\left(\tilde{S}_{SE_j} \tilde{S}_{E_{j}E_{j+1}}\right)\tilde{S}_{SE_{1}}, (L\ge2),
\end{eqnarray}

where
\begin{eqnarray}
\tilde{S}_{SE_{j+1}}=
\begin{pmatrix}
r_{se}&0&t_{se}&0\\
0&I_j&0&0\\
-t_{se}&0&r_{se}&0\\
0&0&0&I_{L-j-2}\\
\end{pmatrix},
\end{eqnarray}
and
\begin{eqnarray}
\tilde{S}_{E_{j}E_{j+1}}=
\begin{pmatrix}
I_j&0&0&0\\
0&r_{ee}&t_{ee}&0\\
0&-t_{ee}&r_{ee}&0\\
0&0&0&I_{L-j-2}\\
\end{pmatrix}.
\end{eqnarray}
Similarly, the scattering matrix for the model consisted of only channel $B$ and its environmental modes $E_j^{B}$ can be obtained by altering the transmissivity $t_{ee}\rightarrow -t_{ee}$ and $t_{se}\rightarrow -t_{se}$.

We can derive the characteristic function of mode $C$ and its environmental modes by setting $\vec{\mu} = [0,\cdots,0,\mu_C,\mu_{E^{C}_{1}},\cdots,\mu_{E^{C}_{L-1}}]$ in Eq.(\ref{eq.JointCharacteristicFunction}).The input-output relation for the characteristic function of dissipative channel $C$ can be given as $\chi_{C,E^{C}}^{\text{out}} = \chi_{C,E^{C}}^{\text{in}}(\tilde{\mathbb{S}}_{E}^{-1}\vec{\mu}_{C,E^C})$, where $\vec{\mu}_{C,E^C}=[\mu_C,\mu_{E^{C}_{1}},\cdots,\mu_{E^{C}_{L-1}}]$. According to Eqs.(\ref{eq.CovarianceMatrix})-(\ref{eq.OrderedMoment}), we can obtain the input and output covariance matrices of mode $C$. For the Gaussian channel based on the covariance matrices, the input-output relation of mode $C$ after $L$ steps can be reexpressed as $\sigma^{\text{out,L}}_{C} = \mathcal{E}_L\left[\sigma^{\text{in}}_{C}\right]$. The dynamical map $\mathcal{E}_L$ is always completely positive and trace-preserving (CPT) and can be formally split as,
\begin{eqnarray}
\mathcal{E}_L = \Phi_{L,L-1} \circ \mathcal{E}_{L-1},
\end{eqnarray}
where the symbol $\circ$ indicates the composition of the superoperators. When $\Phi_{L,L-1}$ is CPT for all $L$, the dynamics is divisible and Markovian. Conversely, when $\Phi_{L,L-1}$ is non-CPT for some values of $L$, the dynamics is indivisible and non-Markovian. In our model, the output covariance matrix of mode $C$ takes the following form,
\begin{eqnarray}
\sigma^{\text{out,L}}_{C}={\cal X}_L\sigma^{\text{in}}_{C}{\cal X}_L^{\text{T}}+{\cal Y}_L,
\end{eqnarray}
where ${\cal X}_L$ and ${\cal Y}_L$ are $2\times2$ real matrices. Introduce the matrix
\begin{eqnarray}
\Lambda_L={\cal Y}_{L,L-1}-\frac{i}{2}\Omega+\frac{i}{2}{\cal X}_{L,L-1}\Omega {\cal X}^{\text{T}}_{L,L-1},
\label{matixLambda}
\end{eqnarray}
with ${\cal X}_{L,L-1}={\cal X}_{L}{\cal X}^{-1}_{L-1}$, ${\cal Y}_{L,L-1}={\cal Y}_{L}-{\cal X}_{L,L-1}{\cal Y}_{L-1}{\cal X}^{\text{T}}_{L,L-1}$, and $\Omega=[0,1;-1,0]$ is the single mode symplectic matrix. The forms of matrices of ${\cal X}_{L}$ and ${\cal Y}_{L}$ can be found in the previous work by one of the authors in Ref.\cite{JiasenJin2018}. The CPT is preserved, for when $\Lambda_L\ge0$; The dynamical map is non-CPT, for when $\Lambda_L<0$, which means that all the negative eigenvalues of $\Lambda_L$ contributes to the non-CPT of the one-step evolution dynamical map $\Phi_{L,L-1}$. Along this line, as proposed by Torre {\it et al.} in Ref.\cite{GTorre2015}, the non-Markovianity of the channel basing on the dynamical indivisiblity can be quantified by
\begin{eqnarray}
\mathcal{N}(L)= \ln{\left(\displaystyle\sum_{j=1}^{L-2}\displaystyle\sum_{k=\pm}\frac{|\lambda_{j,k}| - \lambda_{j,k}}{2}\right)},
\end{eqnarray}
where $\lambda_{j,k}$ are the eigenvalues of the matrix $\Lambda_L$. The quantity $\mathcal{N}(L)$ is always positive semi-definite and the dynamics is Markovian when $\mathcal{N}(L)=0$. After some tedious algebra, in our specific model, the eigenvalues of $\Lambda_L$ is calculated by
\begin{eqnarray}
\lambda_{j,\pm} = \ln{\left(\left(X_{E} \pm  \frac{1}{2}\sqrt{|Y_{E}|^2 + 1}\right)\left[1 - \frac{c_{1,1}^2\left(L\right)}{c_{1,1}^2\left(L-1\right)}\right]\right)},
\label{eq.NMequation}
\end{eqnarray}
where $X_E$ and $Y_E$, as given in Eq.(\ref{eq.GaussianState}), are related to the properties of environmental state and $c_{1,1}(L)$ is the element of matrix $\tilde{\mathbb{S}}_{E}(L)$  at the first row and first column.

We can find that once the scattering matrix $\tilde{\mathbb{S}}_{E}(L)$ is constructed and the environmental states are fixed, the eigenvalues $\lambda_{j,\pm}$ are determined and not related to the properties of system initial state. In our CM, the dissipative channels for modes $B$ and $C$ are identical, which means that (i) the scattering matrices for the same collision have the same coefficients of reflectivity; (ii) the environmental states for $B$ and $C$ are the same.  Thus we can simplify the calculation of non-Markovianity by only consider the dissipative channel for mode $C$. The same results apply to the channel of mode $B$.

\begin{figure}[h]
  \includegraphics[width=1\linewidth]{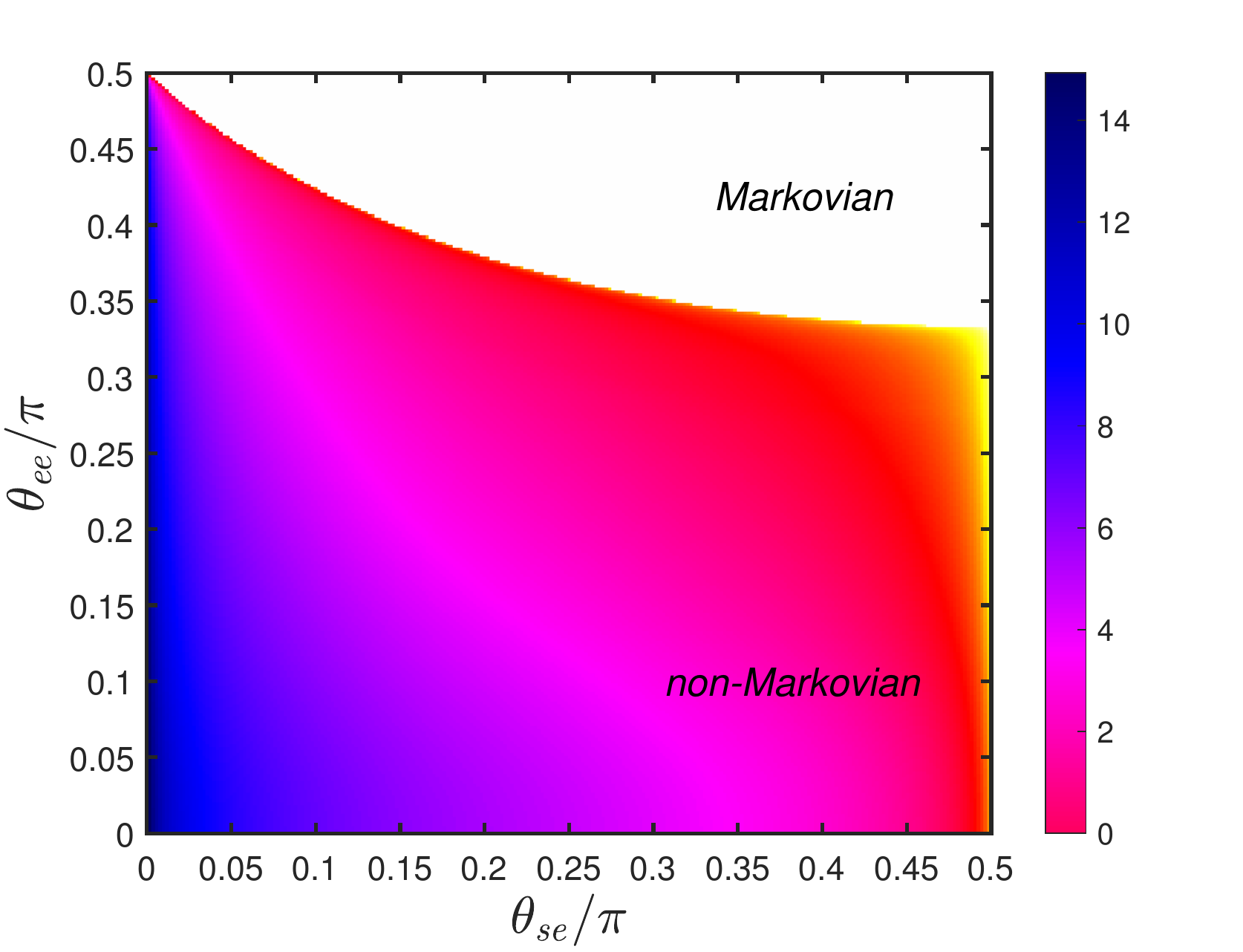}
  \caption{The non-Markovianity in the $\theta_{se}$-$\theta_{ee}$ plane for case of environmental states being vacuum states, with $L = 50$. }
  \label{Fig_NMb}
\end{figure}

In Fig. \ref{Fig_NMb}, we show the non-Markovianity in the $\theta_{ee}$-$\theta_{se}$ plane for vacuum environmental state.
One can see that, by modulating the corresponding transmission angles,  the channel can be tuned from the Markovian to non-Markovian. For $\theta_{ee}/\pi=0.5$, the act of environment-environment collision is a complete reflection, which implies that the information exchange between the old environmental mode and the new environmental mode is forbidden. The old and new environmental modes do not interfere each other. As a consequence, the system can not restore the lost information from the new environmental mode. The dynamics is always Markovian regardless of $\theta_{se}/\pi$. In contrast, the small values of $\theta_{se}$ and $\theta_{ee}$ mean the high transmissivity of BSs, although the information is likely to leak into the environment, the lost information has large probability to flow back into the system via environment-environment collisions.  High transmissivity has a positive effect on improving the degree of non-Markovianity.

In addition, when the collision processes are fixed, which means the scattering matrix is unchanged. The boundary of Markovian and non-Markovian regions does not change if only replace the vacuum environmental state with the generic Gaussian states. The degree of the non-Markovianity of the generic Gaussian states can be calculated as the following \cite{JiasenJin2018},
\begin{eqnarray}
\mathcal{N}_{G}\left(L\right) = \left(2n_E + 1\right)\cosh\left(2r_E\right)\mathcal{N}_{\text{vac}}\left(L\right),
\label{eq.18}
\end{eqnarray}
where $\mathcal{N}_{\text{vac}}$ is  non-Markovianity in the case of the vacuum state as the environmental state. $\mathcal{N}_{G}\left(L\right)$ is zero 
 if and only if $\mathcal{N}_{\text{vac}}\left(L\right)=0$. This supports our previous conclusions: the boundary of Markovian and non-Markovian regions did not change.

\subsection{Tripartite mutual information}\label{TMI}

In order to witness the delocalization of information during the time-evolution of the state of system, we adopt the TMI as the indicator of information scrambling. For our specific model, the TMI regarding to the auxiliary and system modes is defined as
\begin{eqnarray}
I_3(A:B:C) = I_2(A:B) + I_2(A:C) - I_2(A:BC).
\label{eq_TMI}
\end{eqnarray}
In the right-hand side of Eq. (\ref{eq_TMI}), the quantity $I_2$ is the bipartite mutual information (BMI) between the auxiliary mode and the system modes. The BMI quantifies the total correlations between two partitions and whose expression of BMI is defined as follows,
\begin{eqnarray}
I_2(A:X) = S(\rho_{A}) + S(\rho_{X}) - S(\rho_{AX}),
\label{eq_BMI}
\end{eqnarray}
with $X = B, C$ and $BC$. Here $S(\rho_X)=-\text{tr}[\rho_{X}\ln{\rho_X}]$ is the von Neumann entropy of the reduced density matrix for mode $X$.
Alternatively, the von Neumann entropy for a single-mode Gaussian state $\rho$ can be obtained as
\begin{eqnarray}
S(\rho)=\sum_{k=1}^N{f(\nu_k)},
\end{eqnarray}
where $f(x)=\left(x+\frac{1}{2}\right)\ln{\left(x+\frac{1}{2}\right)}-\left(x-\frac{1}{2}\right)\ln{\left(x-\frac{1}{2}\right)}$ and $\nu_{k}$ are the symplectic eigenvalues of the covariance matrix associated to $\rho$ \cite{ASerafini2004,ASerafini2006}.

The negative value of TMI is a diagnostic of quantum information scrambling. This can be understood as the following. According to Eq. (\ref{eq_TMI}), the negative TMI means that the correlation between the auxiliary mode and the joint system modes is more than the sum of correlations between the auxiliary mode to each system mode, namely, the global information about $A$ embedded in the joint system cannot be accessed by locally measurements on each system mode.

\section{Results and discussions}
\label{resultsanddiscussion}
In this section, we will discuss the dynamics of the information in the system modes which interact with their environmental modes. Before that we would consider the dynamics of the joint auxiliary and system modes in the absence of environment. In this case, the joint system (consisted of modes $A$, $B$ and $C$) is closed and the TMI is always zero during the dynamics governed by the interactions between the system modes. This implies the amount of information about $A$ encoded in the joint part $BC$ is equivalent to those in local parts $B$ and $C$. However, when the system-environment interactions are switched on, the interplay between the unitary evolution and the dissipation will lead to rich phenomena in the dynamics.

\subsection{Markovian case}
In this section, we focus on the Markovian dynamics of TMI. We consider the case that the environmental modes $B$ and $C$ are identical. The environmental modes are prepared in the following three types,
\begin{enumerate}
\item Vacuum state: $n_{E_j^{B(C)}} = 0$, $r_{E_j^{B(C)}} = 0$ and $\alpha_{E_j^{B(C)}} = 0$;
\item Squeezed-same state: Squeezed vacuum states with identical squeezing angles $n_{E_j^{B(C)}} = 0$, $r_{E_j^{B(C)}}\ne 0$, $\phi_{E_j^{B(C)}}$ = {\it const.};
\item Squeezed-alternative state: Squeezed vacuum states with the squeezed directions of neighboring mode being perpendicular to each other $n_{E_j^{B(C)}} = 0$, $r_{E_j^{B(C)}} \ne 0$, $\phi_{E_j^{B(C)}} = \pi$ for odd index $j$ and $\phi_{E_j^{B(C)}} = 0$ for even index $j$.
\end{enumerate}

\begin{figure}[h]
  \includegraphics[width=1\linewidth]{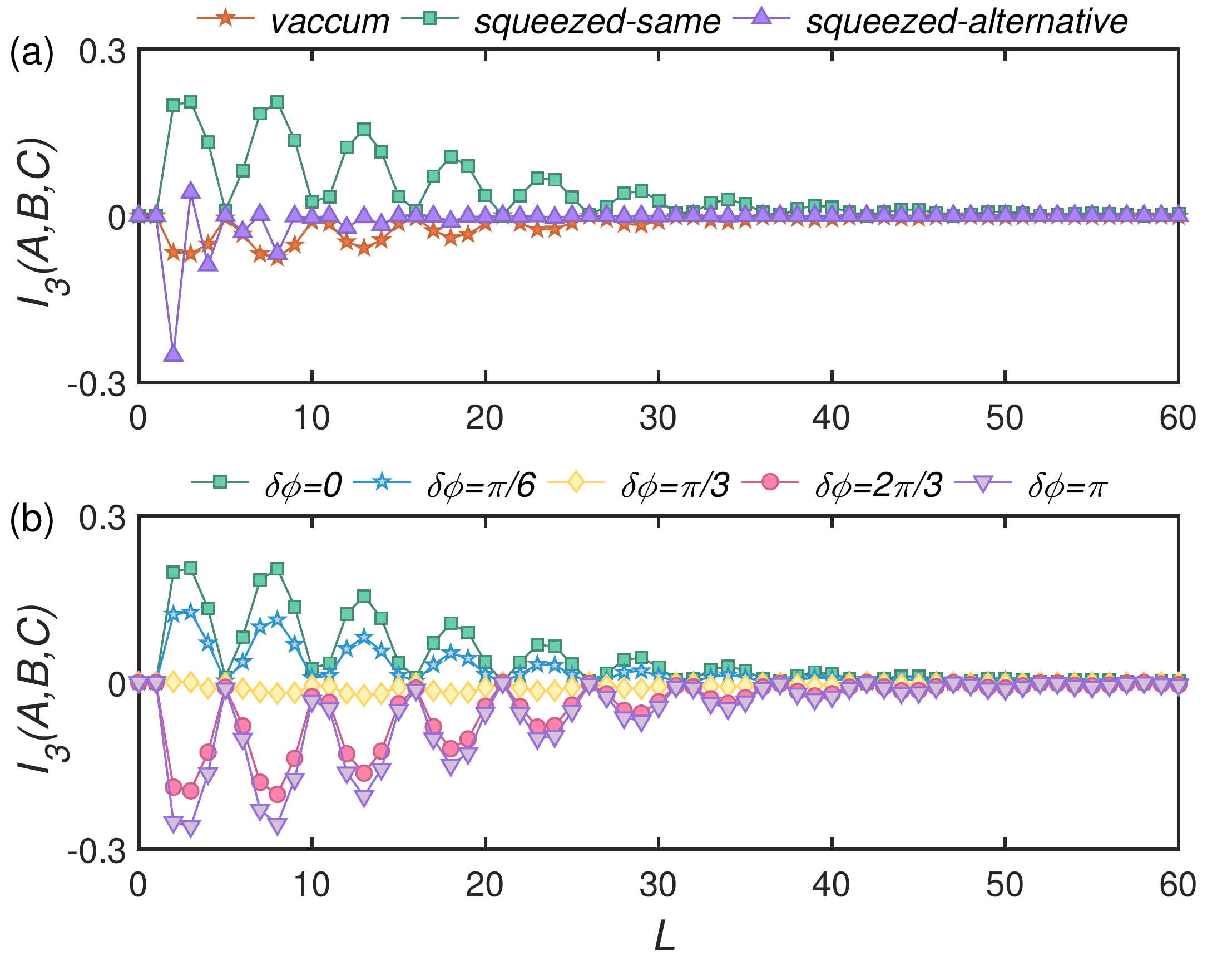}
  \caption{The $L$ dependence of TMI for the various environmental states. Initially, the auxiliary mode is entangled with system mode $B$ via two-mode squeezed state with $\xi_{AB}=1$, the system is in squeezed vacuum state with $\xi_C=1$, the tunable parameters of the BSs are $\theta_{ss} = 0.4\pi$ and $\theta_{se}^{B} = \theta_{se}^{C}= \theta_{ee}^{B} = \theta_{ee}^{C} = 0.35\pi$. (a) The environmental modes are in vacuum state, squeezed-same state with all squeezing angle are $\phi_{E_j^{B(C)}}=0$, $\forall j$ and squeezed-alternative states with perpendicular squeezing direction between neighbors, $\phi_{E_j^{B(C)}} = \pi$ for odd index $j$ and $\phi_{E_j^{B(C)}} = 0$ for even index $j$. The squeezing strengths of the environmental parts are $r_{E_{j}^{B(C)}}=0.5$. (b) For squeezed-same environmental states, the properties of TMI change with phase difference $\delta\phi= \phi_{E_{j}^{B(C)}}-\phi_C$. The squeezing strengths of the environmental parts are $r_{E_{j}^{B(C)}}=0.5$.}
  \label{Fig_initial_state}
\end{figure}

Without loss of generality, we set the parameter of BSs between modes $B$ and $C$ to be $\theta_{ss}$ = $0.4\pi$.
In Fig. \ref{Fig_initial_state}, we show the dynamics of the TMI. The parameters of BSs are set to be $\theta_{se}^{B(C)}$ = $\theta_{ee}^{B(C)}$ = $0.35\pi$ to achieve a Markovian channel. One can find that, for any choice of environmental state, the TMI always asymptotically approaches to zero in the long-time limit (at large $L$), which implies that the information will eventually be lost into the environment. At the early stage of the evolution, however, the TMI shows damping oscillations. In particular, for the vacuum and squeezed-alternative states, the TMI become negative implying the presence of information scrambling. This can be interpreted as the input vacuum and squeezed states or squeezed states with perpendicular squeezing directions become entangled by passing through the BS.

\begin{figure}[h]
	 \includegraphics[width=1\linewidth]{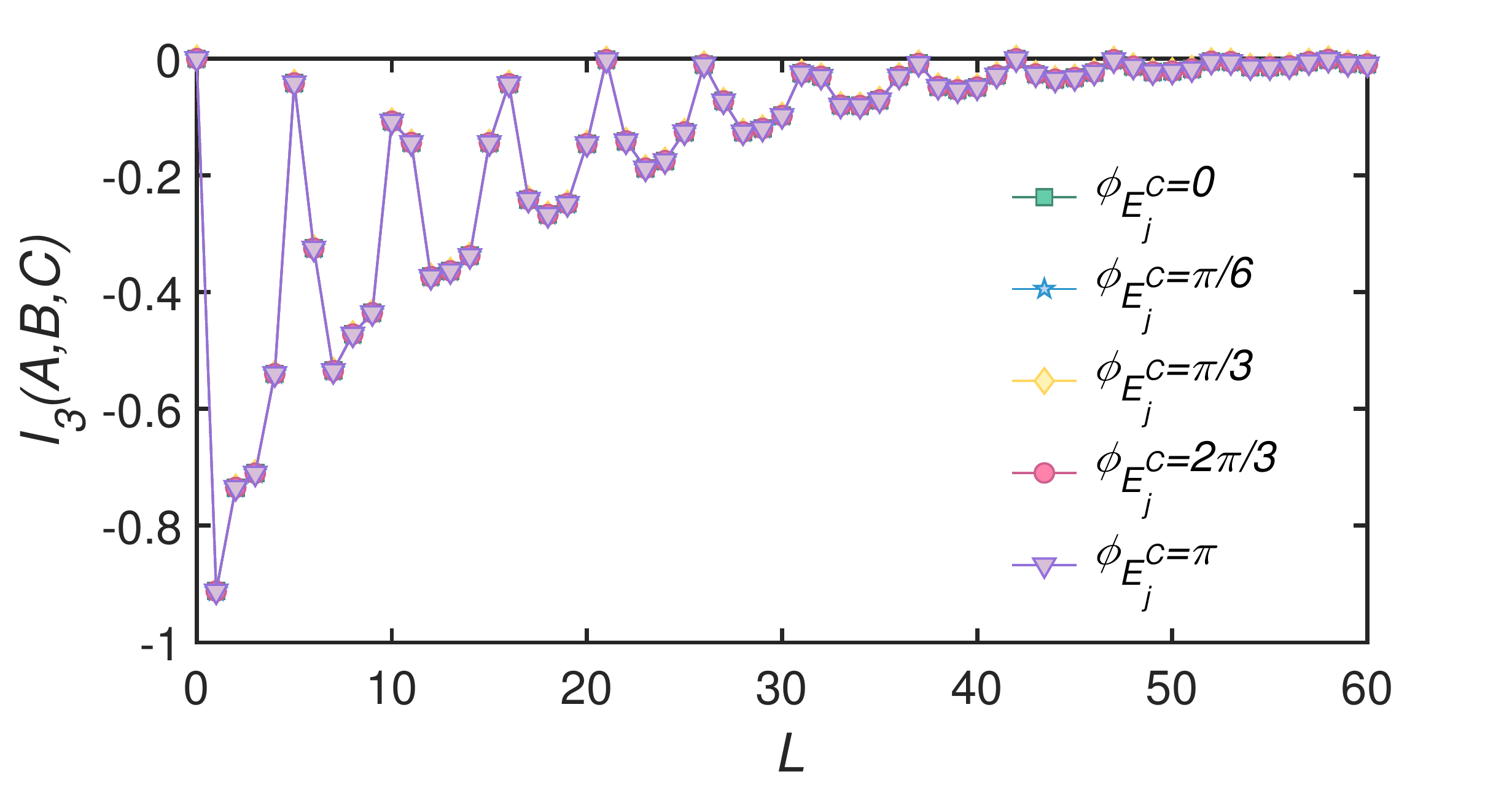}
	 \caption{ The $L$ dependence of TMI for squeezed-same environmental states with different squeezing angles. The system mode $C$ is in the thermal state with $n_C=\sinh^2{\xi_{AB}}$. The squeezing parameters of system are $\xi_{AB}=1$ and the squeezing strengths of environment are $r_{E_j^{B(C)}} = 0.5$. The transmission angles of collisions are $\theta_{ss}=0.4\pi$, $\theta_{se}^{B(C)} = \theta_{ee}^{B(C)} = 0.35\pi$.}
	 \label{Fig_change_angle}
\end{figure}

From Fig. \ref{Fig_initial_state}(a), the squeezed-same environmental state seems to always prevent the information being delocalized. Remind that, for this case, mode $C$ is prepared in the squeezed vacuum state with the same squeezing angle to its environment.
In order to investigate the origin of the information scrambling, we investigate the effect of different squeezing angles between mode $C$ and environmental modes. We define the difference as $\delta \phi = \phi_{E_{j}^{B(C)}}-\phi_C$ and show the time-evolution of TMI for different $\delta\phi$ in Fig. \ref{Fig_initial_state}(b). One can find that as the $\delta\phi$ increasing there is a crossover from the all positive to negative transient TMI during the time evolution. The minimum negative transient TMI appears when the angle difference is $\delta \phi=\pi$, i.e. the squeezing angles of mode $C$ and environments are perpendicular to each other.

So far we have discussed the case that mode $C$ is a squeezed state. On the other hand, since mode $B$ is entangled with the auxiliary mode $A$ in TMSV state, the reduced state of mode $B$ is a thermal state with effective photon number being $\sinh^2{\xi_{AB}}$. We wonder if the unbalanced states in system modes affect the appearance of information scrambling. With this goal in mind, we set mode $C$ to be a thermal state that is identical to the reduced state of mode $B$ and investigate the time-evolution of TMI. The result is shown in Fig.
\ref{Fig_change_angle}. One can see that even if the states of modes $B$ and $C$ are effectively identical to each other and will not entangled through the BS, the transient TMI may be negative during the evolution implying that dissipation is responsible for the information scrambling. In addition, we find that squeezing angle of environmental states does not change the value of TMI.

\subsection{Non-Markovian case}

In this section, we will discuss the effect of non-Markovianity of the channels on information scrambling. We explore the difference of the information dynamics in the non-Markovian case. As mentioned in Sec. \ref{NM}, the non-Markovianity of the channels can be switched on by tuning the parameters $\theta_{se}$ and $\theta_{ee}$ of BSs. Once the parameter $\theta_{se}$ is fixed, the degree of non-Markovianity will decrease with $\theta_{ee}$  and vice versa. Here we set the system mode $C$ to be in a squeezed vacuum state and the environmental modes in the vacuum state.

\begin{figure}[h]
  \includegraphics[width=1\linewidth]{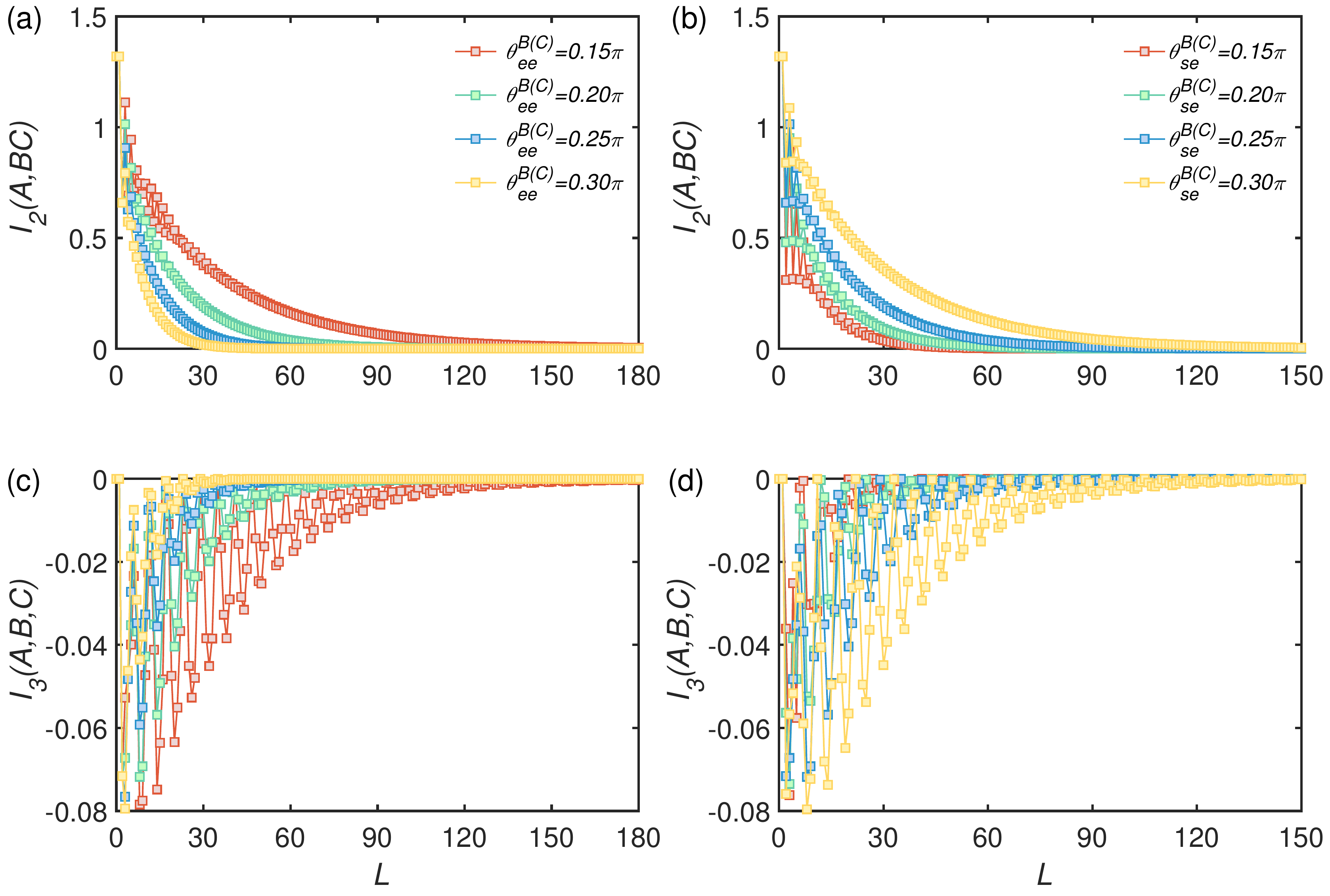}
  \caption{The $L$ dependence of BMI and TMI for the states of the dissipative channel being the vacuum state with $\theta_{ss} = 0.4\pi$. (a) and (c) is the dynamics of BMI and TMI with different transmission angle $\theta_{ee}^{B(C)}$ when $\theta_{se}^{B(C)}=0.25\pi$. (b) and (d) is the dynamic of BMI and TMI with different $\theta_{se}^{B(C)}$ when $\theta_{ee}^{B(C)} = 0.2\pi$.}
  \label{Fig_NM_fixse_vacuum}
\end{figure}

In Fig. \ref{Fig_NM_fixse_vacuum} we show the time-evolution of BMI and TMI for fixed $\theta_{se}$ and $\theta_{ee}$. From Fig. \ref{Fig_NM_fixse_vacuum}(a) and (c), we note that as $\theta_{ee}$ increases the decay of BMI between the auxiliary mode and joint system modes become faster meaning that the encoded information is leaking out into the environment. This is because the larger $\theta_{ee}$ the less information gained by the new environmental mode through environment-environment collision, and consequently less information is flowing back to the system. Nevertheless the information still has the possibility to flow back to the system which is revealed by the nonmonotonic decay of BMI at the early stage of the time evolution. Correspondingly, the TMI also oscillates before the encoded information in the system modes is completely lost.

In contrast, as shown in Fig.\ref{Fig_NM_fixse_vacuum}(b), when the parameter $\theta_{ee}$ is fixed, the decay rate of BMI becomes larger as $\theta_{se}$ decreases. The behavior of TMI shows the similar trends. It is counterintuitive that the strong non-Markovianity speeds up the leaking of information. We interpret this point by considering that non-Markovianity is measured by the indivisibility of dissipative channel. Indeed, the negative eigenvalues become smaller (or say the absolute value become larger) as $\theta_{se}$ decreasing which indicate the strong non-Markovianity, meanwhile, the reflectivity of BS between $C$ and environmental modes is however getting smaller which facilitate the lost of information from system mode.

\begin{figure}[h]
  \includegraphics[width=1\linewidth]{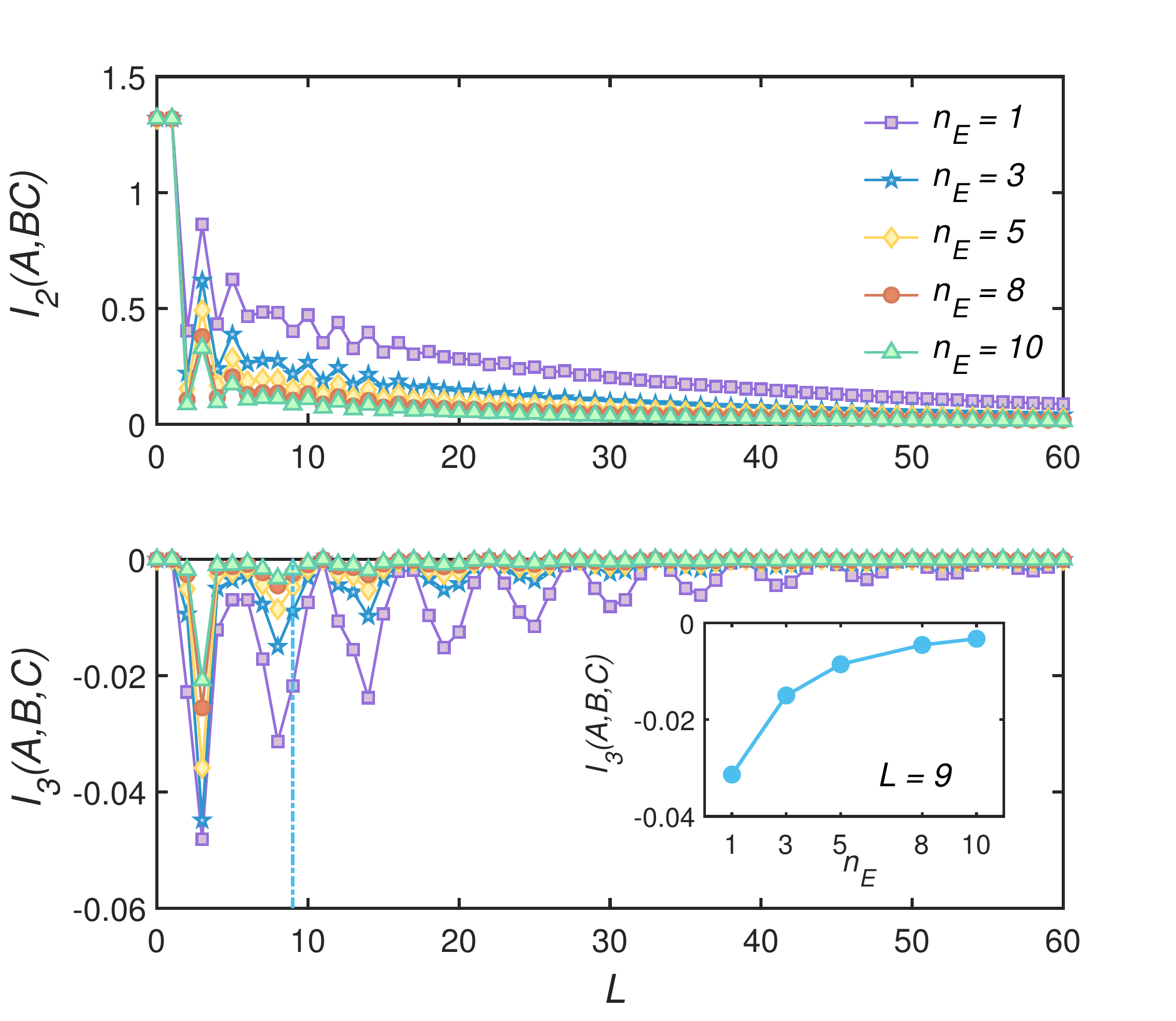}
  \caption{The $L$ dependence of BMI and TMI with the environment part being different thermal states with the thermal mean number $n$ in fixed transmission angle $\theta_{se}^{B(C)}$ = $0.3\pi$ and $\theta_{ee}^{B(C)}$ = $0.15\pi$. Besides, the transmission angle $\theta_{ss}$ = $0.4\pi$.}
  \label{Fig_NM_thermal}
\end{figure}

Another way to tune the degree of non-Markovianity of the channel is to change the states of environmental modes. According to Eq.(\ref{eq.18}), the non-Markovianity will be amplified as the the effective photon number of environmental thermal state increasing. In Fig. \ref{Fig_NM_thermal}, we show the time-evolution of BMI and TMI for different $n_E$. It turns out that the transient value of both BMI and TMI is proportional to the effective photon number as shown in the inset of Fig. \ref{Fig_NM_thermal}(b). In this case, the non-Markovianity only quantitatively modifies the time evolution of information.

We end this section by concluding that the degree of non-Markovianity can indeed affect the dynamics of information, but there is no explicit relationship between presence of non-Markovianity and occurrence of information scrambling in our CM.

\section{Summary}
\label{summary}
In summary, we have proposed an all-optical scheme to simulate the CM. By virtue of the characteristic function formalism, we were able to deal with a large number of bosonic modes and take the interactions between different modes into account. We have considered the cases that system and environmental modes are in various Gaussian states. We have investigated the stroboscopic evolution of information, which is initially encoded the one of the system mode via entangling it with an auxiliary mode, in a three-mode system in the presence of dissipations.

By varying the parameters of BSs in the all-optical network, the dissipative channel can be tuned from Markovian to non-Markovian. In the Markovian case, if system mode is prepared in the squeezed vacuum state, we found that the vacuum and squeezed-alternative environmental states may scramble the information during the dynamics. While the environment being squeezed-same state, there will be a crossover from the absence to presence of information scrambling by changing the difference of squeezing angles between system modes and environmental modes. We have also investigated the case that two system modes are effectively equivalent in terms of thermal states. The results reveal that the occurrence of information scrambling is induced by the dissipation instead of unbalanced system states. In the non-Markovian case, we found that the non-Markovianity can indeed affect the time-evolution of information, however there is not an explicit relationship between non-Markovianity of the channels and appearance of information scrambling in our CM.

Thanks to high stability, arbitrary control of transmissivity, modular nature, and flexible scalability, the all-optical platform can be utilized to simulate the Gaussian boson sampling \cite{hamilton2017,quesada2018,kruse2019,zhong2019}, Anderson localization \cite{crespi2013a}, quantum walk \cite{GeraldiPRL2019}, and quantum state transfer \cite{LouPRL2021,RuPRA2019}. Recently, temporal steering has been utilized as another potential candidate for witnessing information scrambling \cite{LinPRA2021}. The experimental realization of temporal steering in the framework of CM is an intriguing perspective.

\acknowledgments
This work is supported by National Natural Science Foundation of China under Grant No. 11975064.

\section*{APPENDIX}
\label{appendix}

In this appendix, we show the specific forms of Eq.(\ref{eq.ABCCovarianceMatrix}) where the matrix elements $\sigma_{X}$ are the two-dimensional matrices. The detailed forms of the diagonal matrix elements are given as following,
\begin{equation}
 \label{eq_cm_A}
 \sigma_{\text{A}} = \frac{1}{2} \left(
 \begin{array}{cc}
 \cosh(\xi_{AB})&0\\
 0&\cosh(\xi_{AB})\\
 \end{array}
 \right),
 \end{equation}

\begin{equation}
 \label{eq_cm_B}
 \sigma_{B} = \left(
 \begin{array}{cc}
 \alpha^{B}+\beta^{B}&\gamma^{B}\\
 \gamma^{B}&\alpha^{B}-\beta^{B}\\
 \end{array}
 \right),
 \end{equation}

 \begin{equation}
 \label{eq_cm_C}
 \sigma_{C} = \left(
 \begin{array}{cc}
 \alpha^{C}+\beta^{C}&\gamma^{C}\\
 \gamma^{C}&\alpha^{C}-\beta^{C}\\
 \end{array}
 \right),
 \end{equation}
the specific forms of the elements in Eq.(\ref{eq_cm_B}) and Eq.(\ref{eq_cm_C}) are

\begin{equation}
\begin{aligned}
\alpha^{B(C)}&=\frac{1}{2}\left(\cosh (\xi_{AB})  | c_{L,k} |^2 + \cosh (\xi_{C}) |c_{L+2,k}|^2 \right)\\
& + \frac{1}{2} \left[\sum_{w=1}^{L-1} \left(\cosh (2r_{E_{w}^{B}}) + \left(n_{E_{w}^{B}}+ \frac{1}{2}\right) \right)|c_{w,k}|^2 \right]\\
& + \frac{1}{2} \left[\sum_{z=L+3}^{2L+1} \left( \cosh (2r_{E_{z-L-2}^{C}}) + \left(n_{E_{z-L-2}^{C}} + \frac{1}{2}\right) \right)|c_{z,k}|^2 \right]\\
\beta^{B(C)}&=\frac{1}{4}\sinh (\xi_{C}) \left( c_{L+2,k}^{*2} + c_{L+2,k}^2 \right)\\
& + \frac{1}{2} \Re \left( \sum_{w=1}^{L-1} \sinh (2r_{E_{w}^{B}}) e^{i\phi_{E_{w}^{B}}} c_{w,k}^{*2} + \sum_{z=L+3}^{2L+1} \sinh (2r_{E_{z}^{C}}) e^{i\phi_{E_{z-L-2}^{C}}} c_{z,k}^{*2}\right)\\
\gamma^{B(C)}&=\frac{1}{4i} \sinh (\xi_C) \left(c_{L+2,k}^{*2} - c_{L+2,k}^2 \right)\\
& + \frac{1}{2} \Im \left( \sum_{w=1}^{L-1} \sinh (2r_{E_{w}^{B}}) e^{i\phi_{E_{w}^{B}}} c_{w,k}^{*2} + \sum_{z=L+3}^{2L+1} \sinh (2r_{E_{z-L-2}^{C}}) e^{i\phi_{E_{z-L-2}^{C}}} c_{z,k}^{*2}\right)\\
\end{aligned}\nonumber
\end{equation}
where $c_{i,j}$ expresses the matrix element in the $i$-th row and $j$-th column of $\mathbb{S}^{-1}(L)$, and the matrix is given in the main text. The subscript $k$ varies with subscript $B (C)$ of $\alpha^{B(C)}$, $\beta^{B(C)}$ and $\gamma^{B(C)}$, $k=L$ is for subscript $B$ and $k=L+2$ is for subscript $C$. Recall the main text,  $\xi_{AB}$ and $\xi_{C}$ are the squeezed parameters of the joint system $AB$ and system part $C$, respectively. For the environment part, $n_{E_{j}^{B(C)}}$, $r_{E_{j}^{B(C)}}$ and $\phi_{E_{j}^{B(C)}}$ are the thermal mean photon number, squeezed strength and squeezed angle of the $j$-th environmental state in the dissipative channel $B(C)$.

Meanwhile, the off-diagonal matrix elements are governed in the following form,
\begin{equation}
\label{eq_cm_AB}
\sigma_{\text{AB}} = \frac{\sinh (\xi_{AB})}{2} \left(
\begin{array}{cc}
\Re(c_{L,L}^{*})&\Im(c_{L,L}^{*})\\
\Im(c_{L,L}^{*})&-\Re(c_{L,L}^{*})\\
\end{array}
\right),
\end{equation}

\begin{equation}
\label{eq_cm_AC}
\sigma_{\text{AC}} = \frac{\sinh (\xi_{AB})}{2}  \left(
\begin{array}{cc}
\Re(c_{L,L+2}^{*})&\Im(c_{L,L+2}^{*})\\
\Im(c_{L,L+2}^{*})&-\Re(c_{L,L+2}^{*})\\
\end{array}
\right),
\end{equation}

\begin{equation}
\label{eq_cm_BC}
\sigma_{\text{BC}} =\frac{1}{2} \left(
\begin{array}{cc}
\mathrm{K} + \mathrm{M} & \mathrm{P} + \mathrm{Q}\\
\mathrm{P} - \mathrm{Q} & \mathrm{K} + \mathrm{M}\\
\end{array}
\right),
\end{equation}
with the matrix elements in a general form,
 \begin{equation}
 \begin{aligned}
 \mathrm{K}&=\Re\left( \cosh (\xi_{AB})c_{L,L} c_{L,L+2}^{*} + \cosh (\xi_C)c_{L+2,L} c_{L+2,L+2}^* \right) \\
&+ \Re \left[\sum_{w=1}^{L-1} \left( \cosh (2r_{E_{w}^{B}}) + \left(n_{E_{w}^{B}} + \frac{1}{2}\right) \right) c_{w,L} c_{w,L+2}^{*}\right]\\
&+ \Re \left[\sum_{z=L+3}^{2L+1} \left( \cosh (2r_{E_{z-L-2}^{C}}) + \left(n_{E_{z-L-2}^{C}} + \frac{1}{2}\right) \right)c_{z,L} c_{z,L+2}^* \right]\\
 \mathrm{M}&= \sinh (\xi_{C}) \Re \left( c_{L+2,L}^* c_{L+2,L+2}^*\right)\\
 &+\Re \left( \sum_{w=1}^{L-1} \sinh (2r_{E_{w}^{B}}) e^{i\phi_{E_{w}^{B}}} c_{w,L}^*c_{w,L+2}^* + \sum_{z=L+3}^{2L+1} \sinh (2r_{E_{z-L-2}^{C}}) e^{i\phi_{E_{z-L-2}^{C}}} c_{z,L}^*c_{z,L+2}^* \right)  \\
 \mathrm{P}&=\sinh (\xi_{C}) \Im \left( c_{L+2,L}^* c_{L+2,L+2}^*\right)\\
 &+\Im \left( \sum_{w=1}^{L-1} \sinh (2r_{E_{w}^{B}}) e^{i\phi_{E_{w}^{B}}} c_{w,L}^*c_{w,L+2}^* + \sum_{z=L+3}^{2L+1} \sinh (2r_{E_{z-L-2}^{C}}) e^{i\phi_{E_{z-L-2}^{C}}} c_{z,L}^*c_{z,L+2}^* \right)  \\
 \mathrm{Q}&=\frac{1}{2} \Im\left( \cosh (\xi_{AB})c_{L,L} c_{L,L+2}^{*} + \cosh (\xi_C)c_{L+2,L} c_{L+2,L+2}^* \right) \\
 &+ \frac{1}{2} \Im \left[\sum_{w=1}^{L-1} \left( \cosh (2r_{E_{w}^{B}}) + \left(n_{E_{w}^{B}} + \frac{1}{2}\right) \right) c_{w,L} c_{w,L+2}^{*}\right]\\
&+ \frac{1}{2} \Im \left[\sum_{z=L+3}^{2L+1} \left( \cosh (2r_{E_{z-L-2}^{C}}) + \left(n_{E_{z-L-2}^{C}} + \frac{1}{2}\right) \right)c_{z,L} c_{z,L+2}^* \right]\\
 \end{aligned}\nonumber
 \end{equation}

\end{document}